\documentclass{article}
\usepackage{amssymb}
\usepackage{graphicx}

\input{tcilatex}

\begin{document}

\begin{center}
\bigskip

$\mathbf{BELL}$ $\mathbf{NON-LOCALITY}$ $\mathbf{IN}$ $\mathbf{MACROSCOPIC}$ 
$\mathbf{SYSTEMS}$

\bigskip

B. J. Dalton$^{1,2}$

\bigskip

$^{1}$Centre for Quantum and Optical Science, Swinburne University of
Technology, Melbourne, Victoria 3122, Australia

$^{2}$School of Physics and Astronomy, University of Glasgow, Glasgow G12
8QQ, United Kingdom

\bigskip
\end{center}

\textbf{Abstract }The categorization of quantum states for composite systems
as either separable or entangled, or alternatively as Bell local or Bell
non-local states based on local hidden variable theory is reviewed in
Sections 1 and 2, focusing on simple bipartite systems. The significance of
states demonstrating Bell non-locality for settling the long standing
controversy between the Copenhagen interpretation of the quantum measurement
process involving \textquotedblleft collapse of the
wave-function\textquotedblright\ and the alternative interpretation based on
pre-existing hidden variables is emphasized. Although experiments
demonstrating violations of Bell locality in microscopic systems have now
been carried out (see Section 3), there is current interest in finding Bell
non-locality in quantum systems on a macroscopic scale, since this is a
regime where a classical hidden variable theory might still apply. Progress
towards finding macroscopic quantum states that violate Bell inequalities is
reviewed in Section 4.

A new test for Bell non-locality that applies when the sub-system measured
quantities are spin components with large outcomes is described, and applied
to four mode systems of identical massive bosons in Bose-Einstein
condensates.

\section{Copenhagen interpretation and EPR paradox}

To Einstein \cite{Einstein35a}, the Copenhagen quantum interpretation of
what happens in \emph{bipartite systems} when we first measure an observable 
$\Omega _{A}$ in one sub-system $A$ with outcome $\alpha $, and then
immediately measure an observable $\Omega _{B}$ in a second well-separated
sub-system $B$ with outcome $\beta $ seemed counter-intuitive - implying
"instantaneous action at a distance" during the two-step measurement
process. This is known since the 1930's as the \emph{EPR paradox}. According
to the \emph{Copenhagen interpretation}, if the quantum state resulting from
preparation process $c$ is $\widehat{\rho }$, then after the first
measurement the quantum state changes to the conditioned state $\widehat{%
\rho }_{cond}(\alpha |\Omega _{A},\rho )=(\widehat{\Pi }_{\alpha
}^{A}\otimes \widehat{1}^{B})\widehat{\rho }(\widehat{\Pi }_{\alpha
}^{A}\otimes \widehat{1}^{B})/P(\alpha |\Omega _{A},\rho )$, where $\widehat{%
\Pi }_{\alpha }^{A}$ is the projector onto the eigenvector space of $%
\widehat{\Omega }_{A}$ with eigenvalue $\alpha $, and $P(\alpha |\Omega
_{A},\rho )=Tr((\widehat{\Pi }_{\alpha }^{A}\otimes \widehat{1}^{B})\widehat{%
\rho })$ is the probability for state $\widehat{\rho }$ that measurement of $%
\Omega _{A}$ leads to outcome $\alpha $. The density operator is normalized
as $Tr\widehat{\rho }=1$. In general, the reduced density operator $\widehat{%
\rho }_{B}=Tr_{A}\widehat{\rho }$ describing the original state for
sub-system $B$ would be instantly changed to a different state - $Tr_{A}($ $%
\widehat{\rho }_{cond}(\alpha |\Omega _{A},\rho ))$, even though no signal
would have had time to travel between the two well-separated sub-systems.
This effect is referred to as \emph{steering} \cite{Schrodinger35a}. Of
course if $\Omega _{A}$ was immediately measured a second time it is easy to
show that the outcome $\alpha $ would occur with probability $1$. For the
Copenhagenist, the quantum state $\widehat{\rho }$ is not itself a real
object, but only a means of determining the probabilities of the outcomes of
measuring observables - the outcomes being the real objects which are
created by the measurement process on the prepared quantum state. That the
quantum state changes as a result of the measurement of $\Omega _{A}$ with
outcome $\alpha $, merely signifies the probability changing from $Tr((%
\widehat{\Pi }_{\alpha }^{A}\otimes \widehat{1}^{B})\widehat{\rho })\neq 1$
for the original preparation process to $Tr((\widehat{\Pi }_{\alpha
}^{A}\otimes \widehat{1}^{B})\widehat{\rho }_{cond}(\alpha |\Omega _{A},\rho
))=1$ for a new preparation process in which the second part involves
measuring $\Omega _{A}$ with outcome $\alpha $. If we now measure the second
sub-system observable $\Omega _{B}$ the \emph{conditional} \emph{probability}
$P(\beta |\Omega _{B}$ $||\alpha |\Omega _{A},\rho )$ for outcome $\beta $,
given that measurement of $\Omega _{A}$ in the first sub-system $A$ resulted
in outcome $\alpha $, will now be determined from the conditioned state as $%
P(\beta |\Omega _{B},\rho _{cond}(\alpha |\Omega _{A},\rho ))=Tr((\widehat{1}%
^{A}\otimes \widehat{\Pi }_{\beta }^{B})\widehat{\rho }_{cond}(\alpha
|\Omega _{A},\rho ))=Tr(\widehat{\Pi }_{\alpha }^{A}\otimes \widehat{\Pi }%
_{\beta }^{B})\widehat{\rho })/P(\alpha |\Omega _{A},\rho )$. In general
this will be different from the probability $P(\beta |\Omega _{B},\rho )=Tr((%
\widehat{1}^{A}\otimes \widehat{\Pi }_{\beta }^{B})\widehat{\rho })=Tr_{B}((%
\widehat{\Pi }_{\beta }^{B})\widehat{\rho }_{B})$ of outcome $\beta $
resulting from measurement of observable $\Omega _{B}$ for the original
state $\widehat{\rho }$ when no measurement of $\Omega _{A}$ is made.
However, using Bayes' theorem the joint probability for outcomes $\alpha $
for $\Omega _{A}$ and $\beta $ for $\Omega _{B}$ can be determined as $%
P(\alpha ,\beta |\Omega _{A},\Omega _{B},\rho )=P(\beta |\Omega _{B},\rho
_{cond}(\alpha |\Omega _{A},\rho ))\times P(\alpha |\Omega _{A},\rho )=Tr((%
\widehat{\Pi }_{\alpha }^{A}\otimes \widehat{\Pi }_{\beta }^{B})\widehat{%
\rho })$. This is the standard Copenhagen expression for the joint
measurement probability for the measurement of the two observables in the
separated sub-systems if the measurements had been made on the original
state $\widehat{\rho }$ totally independently of each other and in no
particular order. As far as we know, the predictions based on Copenhagen
version of quantum theory are always in accord with experiment. But to
Einstein and others, the Copenhagen theoretical picture was philosophically
unsatisfactory. So the question arose - is it really necessary to invoke the
Copenhagen picture involving the instantaneous change to the quantum state
as a result of the first measurement (the "collapse of the wave function")
to describe what happens, or is there a simpler picture based on classical
probability theory - and involving what we now refer to as hidden variables
- that could also account for all the quantum theory probability predictions?

\subsection{Hidden variable theory and Bell non-locality}

The EPR paradox remained an unresolved issue for many years. However in the
1960's Bell \cite{Bell65a} proposed a quantitative version of a general 
\emph{hidden variable theory} which led to certain inequalities (the \emph{%
Bell inequalities}) involving measurable quantities (such as the mean values
for the measurement outcomes of sub-system observables) which could also be
calculated using standard quantum theory. This suggested that experimental
tests could be carried out to compare the results from quantum theory with
those from hidden variable theory. In hidden variable theory the preparation
process $c$ determines a probabilistic distribution $P(\lambda ,c)$ of \emph{%
hidden variables} $\lambda $. The detailed nature of the hidden variables is
irrelevant, but we require $\dsum\limits_{\lambda }P(\lambda ,c)=1$. The
hidden variables may change with time in accordance with as yet unspecified
dynamical equations, and thus would determine the system's underlying
evolution. Here we just focus on measurements carried out at some particular
time and the hidden variables $\lambda $ are those that apply at the time of
measurement, though they are still determined from the original preparation
process. In accordance with the ideas of classical physics, it may be
assumed that measurements of observables $\Omega $ can be carried out
leading to a possible outcome $\alpha $ without any significant perturbation
of the underlying dynamics. For \emph{bipartite} systems, in each of the two
sub-systems the hidden variables in a \emph{local hidden variable theory}
(LHVT)\ specify separate classical probabilities $P(\alpha |\Omega
_{A},\lambda ,c)$ and $P(\beta |\Omega _{B},\lambda ,c)$ that measurement of
observables $\Omega _{A}$, $\Omega _{B}$ in the respective sub-systems $A,B$
leads to outcomes $\alpha $, $\beta $. The joint probability for outcomes $%
\alpha $ for $\Omega _{A}$ and $\beta $ for $\Omega _{B}$ is then determined
in accord with classical probability theory as $P(\alpha ,\beta |\Omega
_{A},\Omega _{B},c)=\dsum\limits_{\lambda }P(\lambda ,c)P(\alpha |\Omega
_{A},\lambda ,c)P(\beta |\Omega _{B},\lambda ,c)$, and the probability for
outcome $\alpha $ for measuring $\Omega _{A}$ alone would be given by $%
P(\alpha |\Omega _{A},c)=\dsum\limits_{\lambda }P(\lambda ,c)P(\alpha
|\Omega _{A},\lambda ,c)$. This gives the conditional probability for
outcome $\beta $, given that measurement of $\Omega _{A}$ in the first
sub-system $A$ resulted in outcome $\alpha $ as $P(\beta |\Omega _{B}$ $%
||\alpha |\Omega _{A},c)=$ $\dsum\limits_{\lambda }P(\lambda ,c)P(\alpha
|\Omega _{A},\lambda ,c)P(\beta |\Omega _{B},\lambda ,c)$ $%
/\dsum\limits_{\lambda }P(\lambda ,c)P(\alpha |\Omega _{A},\lambda ,c)$.
These expressions may be compared to those from quantum theory. As LHVT
theory is intended to underlie quantum theory, the point is that both the
joint probability $P(\alpha ,\beta |\Omega _{A},\Omega _{B},c)$ and single
probabilities such as $P(\alpha |\Omega _{A},c)$ can be determined from the
LHVT probabilities $P(\lambda ,c),P(\alpha |\Omega _{A},\lambda ,c)$ and $%
P(\beta |\Omega _{B},\lambda ,c)$ \emph{without} requiring a knowledge of
the system density operator $\widehat{\rho }$. States that can be described
via LHVT are referred to as \emph{Bell local} - those that cannot be so
described are \emph{Bell non-local}. However, apart from the differing forms
of the probability expressions, there is a fundamental difference in the
description of what happens in the measurement process. In hidden variable
theory the hidden variables are determined (at least probabilisticaly) in
the preparation process and are carried over to both sub-systems
irrespective of how well they are separated. They then determine the
probabilities for the outcomes $\alpha ,\beta $ of measurements for $\Omega
_{A}$ and $\Omega _{B}$ on the two sub-systems. As we are considering
measurements at the same time, in local hidden variable theory the outcome
of measurement on one sub-system could\emph{\ not} affect that for the other
sub-system. Unlike the Copenhagen theory change to the quantum state as a
result of first measuring $\Omega _{A}$, no instantaneous changes to the
hidden variables is invoked, certainly no change dependent on the outcome $%
\alpha $. Hence, if an experiment could be carried out whose results are in
accord with quantum theory but not in accord with this general hidden
variable theory, the interpretation that quantum theory is under-pinned by a
classical probability theory involving hidden variables would have to be
rejected. As quantum theory has been confirmed in a wide range of other
experimental situations it would be reasonable to accept its validity
(leaving aside the physics of black holes etc.).This does not necessarily
imply though that the Copenhagen interpretation of what happens in the
measurement process would have to be accepted without further discussion,
since other interpretations of quantum theory exist such as the many-worlds 
\cite{Many Worlds} or the Bohmian nonlocal realistic \cite{Bohmian}
interpretations. As these are just different interpretations of quantum
theory no experimental test rules these out. However, the many-worlds
interpretation invokes the idea that every possible measured outcome occurs
with some probability in a separate non-communicating world, and that
separate worlds are created whenever a measurement is made. Philosophically,
this interpretation fails the test of simplicity, which favours the
Copenhagen interpretation based on a single ongoing probabilistic world.
Similar considerations apply to the complicated Bohmian approach, which in
its simplest version involves deterministic particle positions, whose
dynamical evolution depends on the wave-function as determined from the
time-dependent Schrodinger equation, but which can account for experimental
results if Born's probability rule is assumed. Thus, if Bell non-local
states could be found this would resolve the philosophical issue of what
happens in the measurement process in favour of the Copenhagen
interpretation if Occum's razor rules out these alternative interpretations.
There would therefore be quantum states with correlations for the joint
measurement outcomes in separated sub-systems as given by the quantum
expression, which are not accounted for via the classical correlations that
apply to the hidden variable theory expression. Such quantum correlations
are referred to as \emph{Bell correlations}.

\subsection{Quantum and hidden variable theory predictions}

Comparisons between the Copenhagen quantum and local hidden variable theory
predictions can be made based on Bell inequalities involving the\emph{\ mean
values} of the measurement results as well as those based directly on the 
\emph{joint measurement probabilities}. The quantum theory and the LHVT
expressions for the probabilities of joint measurement outcomes $\alpha
,\beta $ for $\Omega _{A},\Omega _{B}$ are 
\begin{eqnarray}
P(\alpha ,\beta |\Omega _{A},\Omega _{B},\rho )_{Q} &=&Tr((\widehat{\Pi }%
_{\alpha }^{A}\otimes \widehat{\Pi }_{\beta }^{B})\widehat{\rho })
\label{Eq.QThyJointProb} \\
P(\alpha ,\beta |\Omega _{A},\Omega _{B},c)_{LHVT} &=&\dsum\limits_{\lambda
}P(\lambda ,c)P(\alpha |\Omega _{A},\lambda ,c)P(\beta |\Omega _{B},\lambda
,c)  \label{Eq.LHVTJointProb}
\end{eqnarray}%
and we then find that the quantum theory and LHVT expressions for the mean
values of the joint measurement outcomes for $\Omega _{A},\Omega _{B}$ are 
\begin{eqnarray}
\left\langle \Omega _{A}\otimes \Omega _{B}\right\rangle _{Q} &=&Tr(\widehat{%
\Omega }_{A}\otimes \widehat{\Omega }_{B})\widehat{\rho }
\label{Eq.QThyMean} \\
\left\langle \Omega _{A}\otimes \Omega _{B}\right\rangle _{LHVT}
&=&\dsum\limits_{\lambda }P(\lambda ,c)\left\langle \Omega _{A}(\lambda
,c)\right\rangle \left\langle \Omega _{B}(\lambda ,c)\right\rangle
\label{Eq.LHVTMean}
\end{eqnarray}%
where $\left\langle \Omega _{A}(\lambda ,c)\right\rangle
=\dsum\limits_{\alpha }\alpha P(\alpha |\Omega _{A},\lambda ,c)$ is the
hidden variable theory mean value for measurement of $\Omega _{A}$ when the
hidden variables are $\lambda $, with a similar result for $\left\langle
\Omega _{B}(\lambda ,c)\right\rangle $. In addition, comparisons can be made
based on measurement outcomes over\emph{\ restricted ranges}. For example if
both $\alpha $ and $\beta $ were restricted to be positive, then the quantum
and LHVT expressions for the joint probabilities of these positive
measurement outcomes are 
\begin{eqnarray}
P(+,+|\Omega _{A},\Omega _{B},\rho )_{Q} &=&Tr((\widehat{\Pi }%
_{+}^{A}\otimes \widehat{\Pi }_{+}^{B})\widehat{\rho })
\label{Eq.QThyJointProbPositiveMeast} \\
P(+,+|\Omega _{A},\Omega _{B},c)_{LHVT} &=&\dsum\limits_{\lambda }P(\lambda
,c)P(+|\Omega _{A},\lambda ,c)P(+|\Omega _{B},\lambda ,c)
\label{Eq.LHVTJointProbPositiveMeast}
\end{eqnarray}%
where $\widehat{\Pi }_{+}^{A}=\dsum\limits_{\alpha >0}$ $\widehat{\Pi }%
_{\alpha }^{A}$ and $P(+|\Omega _{A},\lambda ,c)=\dsum\limits_{\alpha
>0}P(\alpha |\Omega _{A},\lambda ,c)$ are projectors and LHVT probabilities
for positive outcomes for the measurement of $\Omega _{A}$, with similar
expressions for $\Omega _{B}$. Although for simplicity the preceding
discussion has focused on bipartite systems, its generalisation to \emph{%
multipartite} systems is straight-forward.

Note that if the \emph{Heisenberg uncertainty principle} is to be satisfied
in LHVT\ for the case of non-commuting observables, extra contraints would
be required for the sub-system probabilities. Thus for non-commuting quantum
operators where $[\widehat{\Omega }_{A1},\widehat{\Omega }_{A2}]=i\widehat{M}
$, the corresponding LHVT probabilities $P(\alpha _{1}|\Omega _{A1},\lambda
,c),P(\alpha _{2}|\Omega _{A2},\lambda ,c)$ must lead to the required
condition on the LHVT\ standard deviations, namely $\Delta \Omega
_{A1}\times \Delta \Omega _{A1}\geq |\left\langle M\right\rangle |/2$.,
where $(\Delta \Omega _{A1})^{2}=\dsum\limits_{\lambda }\dsum\limits_{\alpha
_{1}}P(\lambda ,c)P(\alpha _{1}|\Omega _{A1},\lambda ,c)(\alpha
_{1}-\left\langle \Omega _{A1}\right\rangle )^{2}$, $\left\langle \Omega
_{A1}\right\rangle =\dsum\limits_{\lambda }\dsum\limits_{\alpha
_{1}}P(\lambda ,c)P(\alpha _{1}|\Omega _{A1},\lambda ,c)\alpha _{1}$, etc.

\section{Categorizing bipartite states}

\subsection{Local hidden states - EPR steering and Bell non-locality}

As explained below, the LHVT sub-system probabilities of measurement
outcomes for sub-system observables may \emph{also} be given by quantum
expressions involving density operators for the \emph{separate} sub-systems
(and not determined from the overall system density operator), and that the
preparation process may also determine probabilities for particular
sub-system quantum density operators to apply. This involves the concept of
local hidden states, which arose first in the case of separable states. Even
for the simple case of bipartite systems, this leads to three different
categories of Bell local states, together with a fourth category in which
LHVT does not apply.

A first question is whether the results for any quantum states describing
two sub-systems can be also described by local hidden variable theory. One
whole class of states that can be so-described are the \emph{separable states%
} \cite{Werner89a}, where the density operator is of the form $\widehat{\rho 
}_{sep}=\dsum\limits_{R}P_{R}\widehat{\rho }_{R}^{A}\otimes \widehat{\rho }%
_{R}^{B}$. Here the preparation process involves preparing each separate
sub-system in states $\widehat{\rho }_{R}^{A}$ and $\widehat{\rho }_{R}^{B}$%
, where $P_{R}$ is the probability that a particular choice $R$ has been
made. Note that for separable states the reduced density operator for each
sub-system $C$ is given by $\widehat{\rho }_{C}=\dsum\limits_{R}P_{R}%
\widehat{\rho }_{R}^{C}$, which in general differs from the sub-system
states $\widehat{\rho }_{R}^{C}$. States where $\widehat{\rho }\neq \widehat{%
\rho }_{sep}$ are the non-separable or \emph{entangled} states. For
separable states the quantum joint probability is given by $P(\alpha ,\beta
|\Omega _{A},\Omega _{B},\rho _{sep})=\dsum\limits_{R}P_{R}P(\alpha |\Omega
_{A},R)P(\beta |\Omega _{B},R)$ where $P(\alpha |\Omega _{A},R)=Tr_{A}((%
\widehat{\Pi }_{\alpha }^{A})\widehat{\rho }_{R}^{A})$ and $P(\beta |\Omega
_{B},R)=Tr_{B}((\widehat{\Pi }_{\beta }^{B})\widehat{\rho }_{R}^{B})$ are
probabilities for the separate sub-system measurement outcomes, which are
given by quantum theory expressions. However, these results are of the same
form as in local hidden variable theory, with the choice $R$ being regarded
as a hidden variable and with $P_{R}\rightarrow P(\lambda ,c)$, $P(\alpha
|\Omega _{A},R)\rightarrow P(\alpha |\Omega _{A},\lambda ,c)$ etc. So as the
separable states can all be given a local hidden variable theory
interpretation, it follows that any state that cannot be so interpreted 
\emph{must} be an entangled state. However, Werner \cite{Werner89a} showed
that there were \emph{some} entangled states that could be interpreted in
terms of local hidden variable theory. Particular examples were the
so-called Werner states \cite{Werner89a}, which are mixed states specified
by a single parameter and involve two sub-systems with equal dimensionality.
This means that the division of quantum states into separable or entangled
does \emph{not} coincide with their division into Bell local and Bell
non-local.

Wiseman et al \cite{Wiseman07a} introduced the idea of a so-called \emph{%
local hidden quantum state} (LHS) which applied when a particular sub-system 
$A$ was also associated with a quantum density operator $\widehat{\rho }%
^{A}(\lambda ,c)$ specified by the hidden variables $\lambda $, and which
determines the LHVT probability $P(\alpha |\Omega _{A},\lambda ,c)$. The
separable states are characterized by\emph{\ both} sub-systems being
associated with a local hidden quantum state, and are examples of quantum
states that can be also described by LHVT (and referred to as \emph{Category
1} states). Within local hidden variable theory we could also have the
situation where only \emph{one} of the two sub-systems ($B$ say), is
associated with a local hidden quantum state $\widehat{\rho }^{B}(\lambda
,c) $ from which the probability is determined as $P(\beta |\Omega
_{B},\lambda ,c)=Tr_{B}(\widehat{\Pi }_{\beta }^{B}\widehat{\rho }%
^{B}(\lambda ,c))$, whilst for the other sub-system $A$ the probability $%
P(\alpha |\Omega _{A},\lambda ,c)$ is not determined from a local hidden
state (referred to as \emph{Category 2} states). Another Bell local
situation is where \emph{neither} sub-system is associated with a local
hidden quantum state (referred to as \emph{Category 3} states). Both these
last two situations still involve entangled quantum states, whilst also
being described by local hidden variable theory. States where there are no
local hidden states are referred to as \emph{EPR steerable states} \cite%
{Wiseman07a}. They allow for the possibility of choosing the measurement for
observable $\Omega _{A}$ to steer sub-system $B$ such that the outcome for
measuring $\Omega _{B}$ can be chosen in advance. The EPR\ steerable states
are all entangled, and include those that are Bell non-local as well as some
that are Bell local and entangled. They are said to exhibit \emph{EPR
correlations}. Bell non-local\emph{\ }states (where the LHVT expression for
the joint probability is not valid at all - will be referred to as \emph{%
Category 4} states, and exhibit the strongest form of correlation between
the two sub-systems. To find whether a state is Bell non-local requires
showing that a Bell inequality - derived from the basic expression $%
\dsum\limits_{\lambda }P(\lambda ,c)P(\alpha |\Omega _{A},\lambda ,c)P(\beta
|\Omega _{B},\lambda ,c)$ for the joint probability - is violated in
experiment.

\subsection{Two categorizations of states}

Clearly then , the division of the states for bipartite systems into
separable and entangled states does not coincide with the categorization of
the states into Bell local and Bell non-local. The relationship between
these two different schemes is shown in Figure.1. For bipartite systems of
identical massive bosons tests for entanglement are set out in \cite%
{Dalton14a} and tests for EPR steering are presented in \cite{Dalton17a}.

\begin{figure}[tbp]
\caption[ClassificationSchemesMk2]{The Quantum Theory and the Local Hidden
Variable Theory Classification Schemes (QTCS and LHVCS). The two categories
of quantum states in the QTCS are shown in the left column and the two basic
categories of quantum states in the LHVCS are shown in the second left
column. The four more detailed categories of quantum states in the LHVCS are
shown in the third left column, whilst the right two columns lists the
features of the four categories of LHVCS states in both the QTCS and LHVCS
schemes.}
\label{ClassificationSchemesMk2}
\includegraphics[width=12cm]{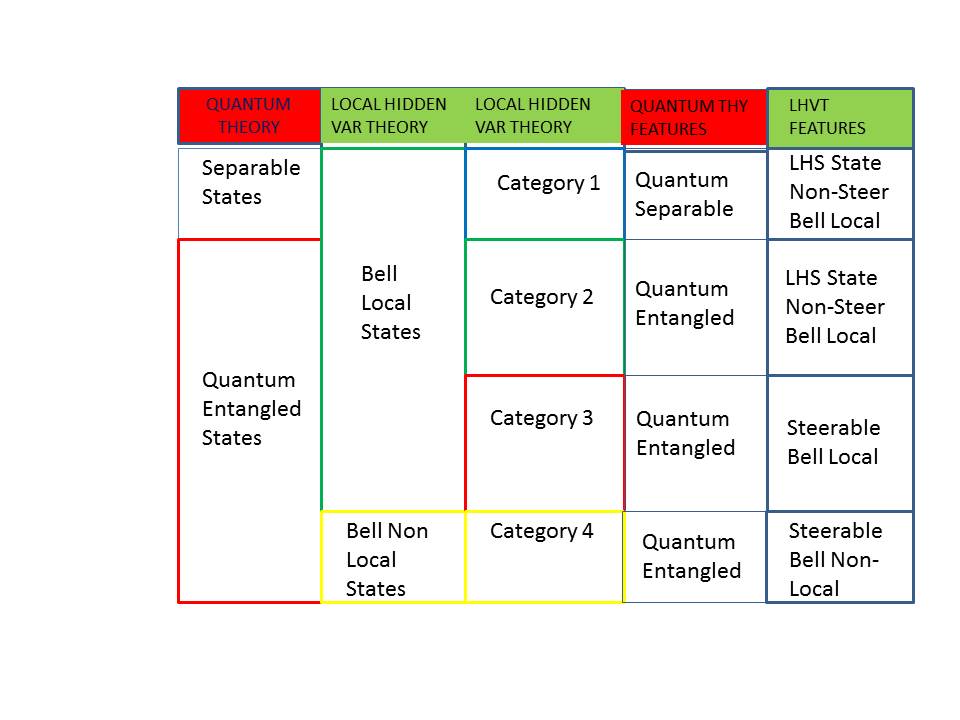}
\end{figure}

%\FRAME{ftbpFUX}{4.0292in}{3.8527in}{0pt}{\Qcb{The Quantum Theory and the
%Local Hidden Variable Theory Classification Schemes (QTCS and LHVCS). The
%two categories of quantum states in the QTCS are shown in the left column
%and the two basic categories of quantum states in the LHVCS are shown in the
%second left column. The four more detailed categories of quantum states in
%the LHVCS are shown in the third left column, whilst the right two columns
%lists the features of the four categories of LHVCS states in both the QTCS
%and LHVCS schemes. }}{}{classificationschemesmk2.jpg}{\special{language
%"Scientific Word";type "GRAPHIC";display "USEDEF";valid_file "F";width
%4.0292in;height 3.8527in;depth 0pt;original-width 9.9998in;original-height
%7.4996in;cropleft "0";croptop "1";cropright "1";cropbottom "0";filename
%'../EPJVersion/ClassificationSchemesMk2.jpg';file-properties "XNPEU";}}

\section{Bell non-locality in microscopic systems}

As pointed out recently \cite{Brunner15a}, there are a multitude of Bell
inequalities that can be derived for both \emph{multi-partite} as well as for%
\emph{\ }bipartite systems, depending on the number of observables
considered in each of the sub-systems and on the number of different
outcomes for each observable. One of the earliest is the famous \emph{CHSH}
Bell inequality for bipartite systems \cite{Clauser69a}. Here there were two
different observables $\Omega _{A1}$, $\Omega _{A2}$ and $\Omega _{B1}$, $%
\Omega _{B2}$ for each sub-system, and measurement of any observable was
restricted to two outcomes - which we choose to be $+1/2$ and $-1/2$. The
CHSH inequality is $|S|\leq 1/2$ , where $S=\left\langle \Omega _{A1}\otimes
\Omega _{B1}\right\rangle +\left\langle \Omega _{A1}\otimes \Omega
_{B2}\right\rangle +\left\langle \Omega _{A2}\otimes \Omega
_{B1}\right\rangle -\left\langle \Omega _{A2}\otimes \Omega
_{B2}\right\rangle $. Suitable physical systems for which this inequality
can be tested include two spin $1/2$ sub-systems, with components of the
spins along various directions being the observables since the measured
outcome is either $+1/2$ or $-1/2$. Another suitable physical system is two
modes of the EM field as the two sub-systems each occupied by one photon,
with the mode polarization being the observable - the outcome being $+1/2$
or $-1/2$ according to whether the outcome is right or left in the case of
circular polarization, or up or across in the case of linear polarization.
These examples are both \emph{microscopic systems}. Experiments testing the
CHSH inequality in microscopic systems have been carried out since the 1970s
(see \cite{Brunner15a} for a recent review), and a violation of the
inequality has now been convincingly demonstrated following numerous
improvements to remove possible loopholes via which the inequality might not
really be violated.

\section{Macroscopic Bell non-locality}

\subsection{Macroscopic systems}

For systems made up of atoms, a system would be regarded as \emph{macroscopic%
} if it contained a very large \emph{number} of atoms and its overall \emph{%
size} scale is large compared to the atomic Bohr radius. Conversely, it
would be\emph{\ }microscopic if the number of atoms was small and its size
was comparable to the Bohr radius. \textquotedblleft
Macroscopic\textquotedblright\ is of course only a qualitative term. Note
that being macroscopic is not necessarily the same as being describable
classically and does not rule out requiring a quantum treatment, though of
course a quantum description is needed for microscopic systems. The main
point of interest is that if \emph{Bell non-locality} is exhibited in a
macroscopic system, then what Einstein regarded as the strangest feature
which distinguishes quantum from classical physics would have occurred in a
situation which ought \emph{not} to require a quantum description. As
discussed in the previous section, Bell non-locality -- which requires
quantum entanglement (even though this is not sufficient to guarantee Bell
non-locality) - has been demonstrated in microscopic systems, but here a
classical theory would be expected to fail, so a Bell inequality violation
is not so surprising. Bell non-locality in a macroscopic system would be
much more unexpected, since this is a regime where a classical theory might
be expected to apply. Bell non-locality requires the quantum state to be
strongly entangled, and entanglement destroying decoherence effects due to
interactions both with the many internal degrees of freedom within a
macroscopic system and due to interactions with the external environment
could be expected to become more prominent for increasingly macroscopic
systems. In comparison, experiments have been carried out with large
molecules (regarded as macroscopic systems) that demonstrate \emph{quantum
interference} between two probability amplitudes, even though quantum
interference effects might be expected not to occur due to decoherence
effects. However, quantum interference is less strange than Bell
non-locality because interference also occurs in classical physics. Showing
that quantum theory is needed for a macroscopic system is always
interesting, but finding Bell locality violations in macroscopic systems
would probably represent the most unusual quantum effect that could be found
-- thus highlighting its importance.

Examples of macroscopic systems in which Bell non-locality could occur
include the following. Optical systems involving large photon numbers in
entangled field modes have been studied as examples of macroscopic systems
even though the notion of system size scale is unclear. A multi-partite
system consisting of a very large of microscopic atomic sub-systems (such as
in cold atomic gases) in which the atomic sub-systems are entangled is
generally be regarded as a potential test bed for macroscopic Bell
non-locality. The quantum effect then involves a macroscopic size scale,
even though the measurement outcomes on the individual sub-systems would be
microscopic. On the other hand, a system in which Bell non-locality occurred
consisting of just two entangled sub-systems (with each containing only a
few modes) would also regarded as demonstrating macroscopic Bell
non-locality if large numbers of particles were associated with each
sub-system. Although the overall system size scale might not be large,
measurement outcomes for each sub-system could have values that are large in
terms of units based on Planck's constant and hence lie in the realm where
classical physics should apply.

Thus, a more significant (though not a requirement) demonstration of
macroscopic Bell non-locality occurs if the physical observables being
measured are those whose outcomes are large in units based on Planck's
constant rather than only having microscopic outcomes. Bell inequality
violations require at least two sub-systems, and although Bell inequalities
have been formulated for multi-partite systems \cite{Brunner15a}, finding a
Bell inequality violation in bipartite macroscopic systems is preferable for
reasons of simplicity as it could involve measurements of a smaller number
of observables. A further consideration is that for systems involving \emph{%
identical massive particles} such as bosonic or fermionic atoms, where the
sub-systems must be defined via distinguishable modes rather than
non-distinguishable atoms, the \emph{symmetrization principle} and the \emph{%
super-selection rules} on particle number are recognised as being important
in regard to tests for quantum entanglement and EPR steering \cite{Dalton14a}%
, \cite{Dalton17a}. Hence physically relevant violations of Bell
inequalities for both microscopic and macroscopic systems would also only
apply for quantum states that comply with the symmetrization principle and
the super-selection rules.

\subsection{Ultra-cold atomic gases and Bell tests}

Although proposals for studying Bell non-locality in macroscopic systems
have been made since the 1980's involving \emph{photonic} systems, systems
made up of a large number of \emph{spin 1/2} parrticles or systems made up
of two\emph{\ high spin} particles, the interest in finding Bell
non-locality in macroscopic systems has grown during the 2000's (see the
review by Reid et al \cite{Reid12a}). This is in part due to experimental
progress in the study of\emph{\ ultracold atomic gases}, which are
macroscopic systems for which a quantum description is required. These
include ultracold bosonic gases, where large numbers of bosonic atoms may
occupy each mode, creating Bose-Einstein condensates. Measurements based on
detecting atom numbers are less error-prone than those involving photon
numbers. For studying bipartite Bell non-locality, two mode systems are
available such as those for bosons with a single spin state trapped in a
double potential well, or for bosons with two different spin states in a
single well. A four mode bipartite system involving two modes associated
with different internal states in each well can also be prepared \cite%
{Riedel10a} using atom-chip techniques. The case treated by Reid et al \cite%
{Reid02a} (see below) applies to this system. Multipartite systems in which
each two state atom is located at a different site on an optical lattice
have also been created \cite{Gross10a}. For ultracold fermionic gases the
situation is not so clear, for although systems with large numbers of
fermionic atoms would be macroscopic, each mode could only be occupied by
fermions with differing spins, and hence many modes would be involved thus
making it difficult to devise bipartite macroscopic systems. In addition to
the experimental progress, a range of theoretical approaches have been found
for deriving Bell inequalities and a large number of different Bell
inequalities have now been obtained. Most only lead to macroscopic
non-locality for multi-partite systems, though a few are associated with
Bell inequality violations for bipartite systems.

We now review some of the Bell inequalities that have been obtained
(presented in historical order) and report on whether experimental tests
have been carried out to find violations of the Bell inequality involved.

\subsection{Mermin (1980), Drummond (1983) Bell inequalities}

There are examples from the 1980s of Bell inequalities applied to
macroscopic systems, though no experimental tests have yet been carried out.
In Ref \cite{Mermin80a} a system consisting of two large spin $s$
sub-systems was considered allowing for measurements of any spin component
to have outcomes from $-s$ to $+s$ in integer steps. For an overall singlet
pure state in which measurement of a spin component in one sub-system leads
to the opposite outcome when the same spin component was measured in the
other, a Bell inequality involving spin components along three unit vectors $%
a,b,c$ of the form $s|\left\langle S_{Aa}\right\rangle -\left\langle
S_{Bb}\right\rangle |\,\geq \left\langle S_{Aa}\otimes S_{Bc}\right\rangle
+\left\langle S_{Ab}\otimes S_{Bc}\right\rangle $ was found. This was found
theoretically to be violated for three distinct coplanar unit vectors, where 
$a,b$ each make an angle $\pi /2+\theta $ with $c$ and hence $\pi -2\theta $
with each other, provided the angle satisfies the condition $0<\sin \theta
<1/2s$. This is a very small range of violating angles if $s$ is large
enough for the system to be considered macroscopic, and the required singlet
state would be difficult to create. Finding particles with large enough $s$
to be macroscopic might possibly be achieved if the "particles": were two
mode BEC with large boson numbers prepared in suitable two mode spin states.

In Ref \cite{Drummond83a} two sub-systems each containing two bosonic modes $%
a_{1},a_{2}$ or $b_{1},b_{2}$ was considered. A maximally entangled state of
the (un-normalized) form $\left( \widehat{a}_{1}^{\dag }\widehat{a}%
_{2}^{\dag }+\widehat{b}_{1}^{\dag }\widehat{b}_{2}^{\dag }\right)
^{N}\left\vert 0\right\rangle $ with a large number of bosons was studied,
and a Bell inequality found involving sub-system boson number-like
observables of the form $(\cos \theta \,\widehat{b}_{1}^{\dag }+\sin \theta
\,\widehat{b}_{2}^{\dag })^{J}(\cos \theta \,\widehat{b}_{1}+\sin \theta \,%
\widehat{b}_{2})^{J}$ for sub-system $B$ with mode annihilation operators $%
\widehat{b}_{1}$, $\widehat{b}_{2}$, with a similar form for sub-system $A$
with mode annihilation operators $\widehat{a}_{1}$, $\widehat{a}_{2}$ -
though here with $\theta =0$. For $J=N\rightarrow \infty $ the inequality is
violated for finite $\theta $ if $3g(\theta )-g(3\theta )-2>0$, where $%
g(\theta )=\exp (-J\theta ^{2}/2)$. Although suitable $\theta $ can be
found, the measurement of the observables for large $J=N$ would be
difficult, requiring the measurement of a very high order quantum
correlation function.

\subsection{MABK (1990-1993) Bell inequalities}

The 1990s saw the introduction \cite{Mermin90a}, \cite{Ardehali92a}, \cite%
{Belinskii93a} of the \emph{MABK} Bell inequalities to treat multipartite
systems with two state spin sub-systems. For GHZ states of the form $%
(\left\vert \uparrow \uparrow \uparrow \uparrow ...\uparrow \right\rangle
+i\left\vert \downarrow \downarrow \downarrow \downarrow ...\downarrow
\right\rangle /\sqrt{2}$ for $n$ sub-systems, a Bell inequality of the form $%
F\leq 2^{n/2}$ (even $n$) is violated for large $n$, where $F=\func{Im}%
(\dsum\limits_{\lambda }P(\lambda )\left\{ 
\begin{array}{c}
(\left\langle S_{x1}(\lambda )\right\rangle +i\left\langle S_{y1}(\lambda
)\right\rangle )....(\left\langle S_{xn}(\lambda )\right\rangle
+i\left\langle S_{yn}(\lambda )\right\rangle ) \\ 
-(\left\langle S_{x1}(\lambda )\right\rangle -i\left\langle S_{y1}(\lambda
)\right\rangle )....(\left\langle S_{xn}(\lambda )\right\rangle
-i\left\langle S_{yn}(\lambda )\right\rangle )%
\end{array}%
\right\} /2i)$ and the spin operators $S_{xi},S_{yi\text{ }}$have outcomes $%
\pm 1$. There is no violation for the bipartite situation $n=2$, which in
any case is microscopic. No experimental tests have yet been carried out,
and the preparation of the GHZ state would be difficult.

\subsection{Reid et al (2002) Bell inequalities}

Around 2000 a Bell inequality originally developed by Clauser et al \cite%
{Clauser78a} was developed by Reid et al \cite{Reid02a} for bipartite
systems in which spin observables of the form $S_{Z}^{A}(2\theta
)=S_{z}^{A}\cos 2\theta +S_{x}^{A}\sin 2\theta $ and $S_{Z}^{B}(2\phi
)=S_{z}^{B}\cos 2\phi +S_{x}^{B}\sin 2\phi $ (both for two mode sub-systems)
were measured and their outcomes $-s_{A}/2,..,+s_{A}/2$ and $%
-s_{B}/2,..,+s_{B}/2$ divided into positive and negative "bins". Although
this would appear to reduce the number of different outcomes to just two for
each sub-system, a situation relevent to macroscopic Bell non-locality still
appears since the Bell inequality is based on considering actual measured
outcomes that are large compared to Planck's constant. The joint
probabilities $P(+,+|S_{Z}^{A}(2\theta ),S_{Z}^{B}(2\phi ))$ and the single
probabilities $P(+|S_{Z}^{A}(2\theta ))$, $P(+|S_{Z}^{B}(2\phi ))$ for
positive outcomes then satisfy a Bell inequality of the form $%
\{P(+,+|\,\theta ,\phi )-P(+,+|\,\theta ,\phi ^{\ast })+P(+,+|\,\theta
^{\ast },\phi )+P(+,+|\,\theta ^{\ast },\phi ^{\ast })\}/\{P(+|\,\theta
^{\ast })+P(+|\,\phi )\}\leq 1$. For the maximally entangled state of the
(un-normalized) form $\left( \widehat{a}_{1}^{\dag }\widehat{a}_{2}^{\dag }+%
\widehat{b}_{1}^{\dag }\widehat{b}_{2}^{\dag }\right) ^{N}\left\vert
0\right\rangle \varpropto \dsum\limits_{m=-s}^{s}\left\vert s,m\right\rangle
_{A}\left\vert s,m\right\rangle _{B}$ (where here $s_{A}=s_{B}=s=N/2$), Bell
inequality violations occurred for a range of parameters $\theta $, $\theta
^{\ast }$, $\phi $, $\phi ^{\ast }$ for both small and large $N$. The large $%
N$ case corresponds to a macroscopic Bell locality violation in a bipartite
system. The original application was to photonic systems, but the theory
also applies for ultracold atomic gases. \ So far, no experimental tests
have been made. As for Ref \cite{Drummond83a} the two mode state would be
difficult to prepare.

\subsection{Collins et al (2002) Bell inequalities}

Also in the early 2000's Collins et al \cite{Collins02a} found a different
approach (\emph{CGLMP}) to deriving Bell inequalities. For bipartite systems
the treatment assumed the existence of HVT probabilities of the form $%
P(\alpha _{j},\alpha _{k},\beta _{l},\beta _{m}|\Omega _{A1},\Omega
_{A2},\Omega _{B1},\Omega _{B2},c)$ (denoted $c_{j,k,l,m}$,) for
simultaneous measurement outcomes $\alpha _{j},\alpha _{k},\beta _{l},\beta
_{m}$ for the pairs of sub-system observables $\Omega _{A1},\Omega
_{A2},\Omega _{B1},\Omega _{B2}$. Clearly, $\dsum%
\limits_{j,k,l,m}c_{j,k,l,m}=1$. The outcomes themselves were the hidden
variables, and the hidden variable theory was stated to be local. Although
this is not stated, LHVT would require the factorization of the
probabilities into $P(\alpha _{j},\alpha _{k}|\Omega _{A1},\Omega _{A2},c)$ (%
$a_{j,k}$ for short) and $P(\beta _{l},\beta _{m}|\Omega _{B1},\Omega
_{B2},c)$ ($b_{l,m}$ for short), thus $c_{j,k,l,m}=a_{j,k}\times b_{l,m}$.
The separate sub-system probabilities would satisfy the constraints $%
\dsum\limits_{j,k}a_{j,k}=1$ and $\dsum\limits_{l,m}b_{l,m}=1$. The
observables for each sub-system were assumed to have the same number of
outcomes (listed as $j,k,l,m=0,1,..d-1\,($mod $d),$- thus $\alpha _{d}\equiv
\alpha _{0}$ etc.). Probabilities for outcomes for one observable for each
sub-system would be obtained as $P(\alpha _{j},\beta _{l}|\Omega
_{A1},\Omega
_{B1},c)=\dsum\limits_{k,m}c_{j,k,l,m}=\dsum\limits_{k,m}a_{j,k}b_{l,m}$
etc., and probabilities for outcomes for one observable of a specific
sub-system given by expressions such as $P(\alpha _{j}|\Omega
_{A1},c)=\dsum\limits_{k,l,m}c_{j,k,l,m}=\dsum\limits_{k}a_{j,k}=A1(j)$ for
short.

The idea behind the \emph{CGLMP }inequalities involves considering joint
outcomes for pairs of observables $\Omega _{A},\Omega _{B}$ for the two
sub-systems in which either the outcomes are for the \emph{same} members of
the two outcome lists or where the outcomes refer to \emph{different}
members of the two lists. Probabilities for the same listed outcomes for
specific observables for the two sub-systems are given by expressions such
as $P(\Omega _{A1}=\Omega
_{B1})=\dsum\limits_{j}\dsum\limits_{k,m}c_{j,k,j,m}=\dsum\limits_{j}\dsum%
\limits_{k,m}a_{j,k}b_{j,m}=\dsum\limits_{j}A1(j)\times B1(j)$, which is the
probability for all outcomes listed $j$ with $\Omega _{A1}$ leading to $%
\alpha _{j}$ and all outcomes for $\Omega _{B1}$ leading to $\beta _{j}$.
Probabilities for outcomes for specific observables for the two sub-systems
where the listed outcomes are shifted are given by expressions such as $%
P(\Omega _{B1}=\Omega
_{A2}+1)=\dsum\limits_{k}\dsum\limits_{j,m}c_{j,k,(k+1),m}=\dsum\limits_{k}%
\dsum\limits_{j,m}a_{j,k}b_{k+1,m}=\dsum\limits_{k}A2(k)\times B1(k+1)$,
where here we consider all outcomes with $\Omega _{A2}$ leading to $\alpha
_{k}$ and all outcomes for $\Omega _{B1}$ leading to $\beta _{k+1}$.
Combinations of such joint probablities for the \emph{four} possible pairs
of observables $\Omega _{A},\Omega _{B}$ then involve the basic LHVT
probabilities $c_{j,k,l,m}=a_{j,k}\times b_{l,m}$, and are then used to
derive Bell inequalities.

For example, combinations of joint measurement probabilities of the form $%
I=P(\Omega _{A1}=\Omega _{B1})+P(\Omega _{B1}=\Omega _{A2}+1)+P(\Omega
_{A2}=\Omega _{B1})+P(\Omega _{B2}=\Omega _{A1})$ were stated to satisfy $%
I\leq 3$ for LHVT. Based just on HVT without assuming locality, we have $%
I=\dsum\limits_{j,k,m}$ $%
(c_{j,k,j,m}+c_{j,k,(k+1),m}+c_{j,k,m,k}+c_{j,k,m,j})$. For a given choice
of $j,k,m$ there is no reason why all four terms cannot be non-zero (in
terms of the notation in Ref \cite{Collins02a}, $r^{^{\prime }}+s^{^{\prime
}}+t^{^{\prime }}+u^{^{\prime }}=0$ for each term). So as $%
\dsum\limits_{j,k,l,m}c_{j,k,l,m}=1$ and each of the four terms is just a
partial contribution to this last equation, it follows that each of the four
terms must be between $0$ and $1$ -since the other part of the contribution
also just involves positive terms. Thus, general HVT would imply that $I\leq
4$, as is stated in Ref. \cite{Collins02a}. Also, if the LHVT condition $%
c_{j,k,l,m}=a_{j,k}\times b_{l,m}$ is invoked we then find that $%
I=\dsum%
\limits_{j=0}^{d-1}(A1(j).B1(j)+A2(j).B1(j)+A1(j).B2(j))+(A2(0).B1(1)+A2(1).B1(2)+....+A2(d-2).B1(d-1)+A2(d-1).B1(0)) 
$. $.$The individual measurement probabilities $A1(j),A2(j),B1(j),B2(j)$ are
of course all positive and satisfy constraints such as $\dsum%
\limits_{j}A1(j)=1$ etc. For LHVT it is stated in Ref. \cite{Collins02a}
that $I\leq 3$, though no proof is given for this result. However, by
multiplying the two constraints for the $A1(j)$ and the $B1(j)$, it is easy
to establish that $\dsum\limits_{j=0}^{d-1}A1(j).B1(j)\leq 1$, since this
expression is a partial positive contribution to the overall product of $1$,
and the other contribution is also positive. Similar arguements show that $%
\dsum\limits_{j=0}^{d-1}A2(j).B1(j)\leq 1$ and $\dsum%
\limits_{j=0}^{d-1}A1(j).B2(j)\leq 1$. Finally, by multiplying the
constraints for $A2(j)$ and $B1(j)$ we see that $%
(A2(0).B1(1)+A2(1).B1(2)+....+A2(d-2).B1(d-1)+A2(d-1).B1(0))\leq 1$, this
expression again being a positive partial contribution to the overall
product of $1.$Adding together these four inequalities gives $I\leq 4$ for
the LHVT case, rather than $I\leq 3$ as stated in Ref. \cite{Collins02a}. A
convincing proof of the $I\leq 3$ result for the LHVT case is needed. Hence
there is now some doubt as to whether this inequality is a general
requirement for LHVT, so its violation does not necessarily show that
quantum theory is required to explain the measurements. Other similar
expressions to $I$ also led to Bell inequalities, but similiar issues apply
in these cases as well.

For the (unnormalized) state $\dsum\limits_{m=-s}^{s}\left\vert
s,m\right\rangle _{A}\left\vert s,m\right\rangle _{B}$ (see above) the
quantum expression for $I$ is found to be greater than $3$ for all $d=2s+1$,
corresponding to a Bell inequality violation in a macroscopic system if $%
s=N/2$ is large. However, this violation involved introducing physical
quantities $\Omega _{A1},\Omega _{A2},\Omega _{B1},\Omega _{B2}$ as
Hermitian operators defined by their eigenvalues and eigenvectors (see Eq.
(13) in Ref \cite{Collins02a}), the latter being linear combinations of the $%
\left\vert s,m\right\rangle _{A(B)}$ .As the operators turn out to be
off-diagonal in these basis states, it is not obvious what physical
observable they correspond to. Finally, no experimental tests of the Bell
inequalities have been carried out, so for several reasons the Collins et al 
\cite{Collins02a} inequalities do not look promising as a vehicle for
finding macroscopic Bell non-locality.

\subsection{CFRD (2007-2011) Bell inequalities}

Around 2010, a further approach in form of the \emph{CFRD} Bell inequalities
was obtained \cite{Cavalcanti07a}, \cite{He10a}, \cite{He11a}. From the
basic LHVT expression (\ref{Eq.LHVTMean}) for mean values a Bell inequality
for bipartite systems with two observables per sub-system is given by $%
\left\langle |(\Omega _{A1}+i\Omega _{A2})(\Omega _{B1}-i\Omega
_{B2})|^{2}\right\rangle \geq |\left\langle (\Omega _{A1}+i\Omega
_{A2})(\Omega _{B1}-i\Omega _{B2})\right\rangle |^{2}$ applies. This may
also be written as $\left\langle (\Omega _{A1}^{2}+\Omega _{A2}^{2})(\Omega
_{B1}^{2}+\Omega _{B2}^{2})\right\rangle \geq |\left\langle (\Omega
_{A1}\Omega _{B1}+\Omega _{A2}\Omega _{B2})\right\rangle +i\left\langle
(\Omega _{A2}\Omega _{B1}-\Omega _{A1}\Omega _{B2})\right\rangle |^{2}$,
where unlike the CHSH inequality an CFRD inequality involves both first
moment $\left\langle \Omega _{Ai}\Omega _{Bj}\right\rangle $ and second
moment $\left\langle \Omega _{Ai}^{2}\Omega _{Bj}^{2}\right\rangle $
correlation functions. CFRD\ type inequalities are also obtained for the
multimode case. For bipartite systems each consisting of a single bosonic
mode $a$ or $b$ the theory has been applied \cite{He10a} for the choice of
quadrature variables $\Omega _{A1}=x_{A}$, $\Omega _{A2}=p_{A}$ and $\Omega
_{B1}=x_{B}$, $\Omega _{B2}=p_{B}$. No Bell inequality violation was found
for the GHZ symmetric state ($\left\vert 0\right\rangle _{A}\left\vert
1\right\rangle _{B}+\left\vert 1\right\rangle _{A}\left\vert 0\right\rangle
_{B})/\sqrt{2}$ - which is microscopic anyway. However, by relating the
quadrature operators to the two mode spin operators (see Ref. \cite%
{Dalton17a}) one can show that to violate the CFRD\ inequality requires
finding a quantum state such that $\left\langle \Delta \widehat{S}%
_{x}^{2}\right\rangle +\left\langle \Delta \widehat{S}_{y}^{2}\right\rangle +%
\frac{1}{4}<0$, showing that it can never be violated for any quantum state.
For bipartite systems each consisting of two bosonic modes $a_{1}$, $a_{2}$
or $b_{1}$, $b_{2}$ the theory has also been applied \cite{He11a} for the
choice of spin variables $\Omega _{A1}=S_{x}^{A}$, $\Omega _{A2}=S_{y}^{A}$
and $\Omega _{B1}=S_{x}^{B}$, $\Omega _{B2}=S_{y}^{B}$.. For the
(unnormalized) quantum state $\dsum\limits_{m=-s}^{s}\,r_{m}\,\left\vert
s,m\right\rangle _{A}\left\vert s,m\right\rangle _{B}$ (where the $r_{m}$
were chosen to optimize the non-locality condition), no violation of the
Bell inequality was found except for the case $s=\frac{1}{2}$, corresponding
to one boson in each subsystem - a microscopic case. Other choices of
observables such as $\Omega _{A1}=\underrightarrow{S}^{A}\cdot 
\underrightarrow{u_{1}}$, $\Omega _{A2}=\underrightarrow{S}^{A}\cdot 
\underrightarrow{u_{2}}$, $\Omega _{B1}=\underrightarrow{S}^{B}\cdot 
\underrightarrow{v_{1}}$, $\Omega _{B2}=\underrightarrow{S}^{B}\cdot 
\underrightarrow{v_{2}}$ and other choices of quantum state could perhaps
result in a Bell inequality violation - however such cases are yet to be
explored. When applied to multi-partite situations, the CFRD inequalities do
lead to Bell inequality violations for any spin $s$ when the numbers of
sub-systems becomes large enough (see Figure in Ref \cite{He11a}). The
reason for this effect is still not understood. So far, no experimental
tests have been made.

\subsection{Tura et al (2014) Bell inequalities}

More recent discussions of Bell non-locality in many-body systems are
presented in Refs \cite{Brunner15a}, \cite{Tura14a}, \cite{Schmied16a} and 
\cite{Engelsen17a}, based on treating the allowed LHVT probabilities in
terms of the theory of polytopes These contain examples of multipartite Bell
inequalities, with applications to systems such as two state atoms located
at different sites in an optical lattice. Here each identical atom $i=1,..,N$
is treated as a distinguishable two mode pseudo-spin sub-system.
Measurements on one of two chosen spin components $M_{i0}$ or $M_{i1}$ for
the $i$th atom sub-system are considered, the two possible outcomes being
designated as $\alpha _{i}=\pm 1$. Defining $S_{0}=\dsum\limits_{i}\left%
\langle M_{i0}\right\rangle $, $S_{00}=\dsum\limits_{i,j(i\neq
j)}\left\langle M_{i0}M_{j0}\right\rangle $, $S_{11}=\dsum\limits_{i,j(i\neq
j)}\left\langle M_{i1}M_{j1}\right\rangle $ and $S_{01}=\dsum\limits_{i,j(i%
\neq j)}\left\langle M_{i0}M_{j1}\right\rangle $ involving the mean values
of single measurements on individual spins or joint measurements on
different spins, a Bell inequality $2S_{0}+S_{01}+2N+(S_{00}+S_{11})/2\geq 0$
has been derived \cite{Tura14a}. Bell inequality violations were predicted
for Dicke states \cite{Dicke54a}. These have the advantage of being the
lowest energy eigenstates for certain many-body Hamiltonians that describe
physical systems, such as $N$ spins interacting via two-body ferromagnetic
coupling, so experimental situations to search for Bell inequality
violations were seen as being readily available. However, Bell correlations
based on this inequality have actually been found \cite{Schmied16a}, \cite%
{Engelsen17a} in systems involving $5\times 10^{2}$ and $5\times 10^{5}$.
bosonic atoms prepared in spin squeezed states. Two component bosonic atoms
were localised on optical or magnetic lattices, with the two spin states
being coupled via Rabi fields. Spin squeezing occured due to inter-atomic
collisions. In these systems the indistinguishability of the identical atoms
and the effect of super-selection rules that rule out sub-system states with
coherences between different boson numbers was ignored, as there is just one
atom in each separated spatial mode on each different lattice site. However,
there is no macroscopic violation of Bell locality in the bipartite case,
since this would only correspond to just two atoms. Nevertheless, these two
experiments provide examples of Bell non-locality in a macroscopic system,
albeit for the multi-partite situation.

\subsection{Dalton (2017) - Generalised CHSH Bell inequality}

Finally, a more standard application of LHVT for bipartite systems each with
two bosonic modes and involving spin observables, leads to the following
Bell inequality - $|S|\leq \frac{1}{2}\left\langle N_{A}\right\rangle
\left\langle N_{B}\right\rangle $, where $S=\left\langle \Omega _{A1}\otimes
\Omega _{B1}\right\rangle +\left\langle \Omega _{A1}\otimes \Omega
_{B2}\right\rangle +\left\langle \Omega _{A2}\otimes \Omega
_{B1}\right\rangle -\left\langle \Omega _{A2}\otimes \Omega
_{B2}\right\rangle $ and $\Omega _{A1}=\underrightarrow{S}^{A}\cdot 
\underrightarrow{u_{1}}$, $\Omega _{A2}=\underrightarrow{S}^{A}\cdot 
\underrightarrow{u_{2}}$, $\Omega _{B1}=\underrightarrow{S}^{B}\cdot 
\underrightarrow{v_{1}}$, $\Omega _{B2}=\underrightarrow{S}^{B}\cdot 
\underrightarrow{v_{2}}$ are components of the spin observables, with $N_{A}$%
, $N_{B\text{ }}$giving the number of bosons in each sub-system. This
inequality is a generalisation of the CHSH\ inequality and its derivation is
similar. Details are given in Ref. \cite{Dalton17a} (see version 1, section
6.1 ). For the case of spin $\frac{1}{2}$ sub-systems this reduces to the
CHSH inequality. Several different quantum states have been tested for
violation of this Bell inequality. These included: (a) the relative phase
eigenstate $\tsum\limits_{k\,=\,-n/2}^{+n/2}\exp (ik\theta )\,\left\vert 
\frac{n}{2},k\right\rangle _{A}\left\vert \frac{n}{2},-k\right\rangle _{B}/%
\sqrt{n+1}$ \cite{Dalton12a} (b) the maximally entangled state $%
\tsum\limits_{k\,=\,-n/2}^{+n/2}\,\left\vert \frac{n}{2},k\right\rangle
_{A}\left\vert \frac{n}{2},k\right\rangle _{B}/\sqrt{n+1}$, (c) the Werner 
\cite{Werner89a} states $\widehat{\rho }_{W}=(d^{3}-d)^{-1}\left( (d-\phi )%
\widehat{1}+(d\phi -1)\widehat{V}\right) $, (where $d=n+1$ and $\widehat{1}$
is the unit operator defined in the $d\times d$ space whose basis vectors
are $\left\vert \frac{n}{2},k\right\rangle _{A}\left\vert \frac{n}{2}%
,l\right\rangle _{B}$ with $k,l=-n/2,-n/2+1,....,+n/2$, and $\widehat{V}$ is
the flip operator defined by $\widehat{V}\left\vert \frac{n}{2}%
,k\right\rangle _{A}\left\vert \frac{n}{2},l\right\rangle _{B}=\left\vert 
\frac{n}{2},l\right\rangle _{A}\left\vert \frac{n}{2},k\right\rangle _{B}$.
Physical restrictions on the parameter $\phi $ \ are $+1\geq \phi \geq -1$)
and (d) the angular momentum eigenstates

$\dsum\limits_{k_{A},k_{B}}$.$C(\frac{N_{A}}{2},\frac{N_{B}}{2}%
,J;k_{A},k_{B},K)\,\left\vert \frac{N_{A}}{2},k_{A}\right\rangle
_{A}\left\vert \frac{N_{B}}{2},k_{B}\right\rangle _{B}$.where the

$C(\frac{N_{A}}{2},\frac{N_{B}}{2},J;k_{A},k_{B},K)$ are Clebsch-Gordon
coefficients. Numerical optimization methods to choose the four spin
components were used. For these four cases the Bell inequality was only
violated occurred for the microscopic case where $N_{A}=N_{B}=1$, which just
corresponds to the CHSH\ situation. Other states, such as spin squeezed
states would be worth studying. As the mean values $\left\langle 
\underrightarrow{S}^{A}\cdot \underrightarrow{u}\otimes \underrightarrow{S}%
^{B}\cdot \underrightarrow{v}\right\rangle $ for products of these spin
operators for bipartite systems (each containing two modes) can be measured
fairly easily using mode couplers with suitable phases and pulse lengths,
then finding a suitable quantum state with large $\left\langle
N_{A}\right\rangle $ and $\left\langle N_{B}\right\rangle $ where the Bell
inequality was violated would provide a case of macroscopic Bell
non-locality.

\section{Conclusions}

A number of different forms of Bell inequalities have been obtained over the
last four decades, which could be tested to find Bell non-locality in
macroscopic systems. A successful outcome would be highly significant,
establishing the priority of the Copenhagen quantum theory over local hidden
variable theories for systems where a classically based theory might be
expected to apply. Up to the present, only two experiments \cite{Schmied16a}%
, \cite{Engelsen17a} have achieved this, based on the versions of Bell
inequalities derived by Tura et al \cite{Tura14a}. These experiments however
are for the multi-partite situation rather than the simpler bipartite case.
The derivation of testable Bell inequalities for macroscopic bipartite
(rather than multi-partite) systems is an ongoing issue, as is the
experimental search for more cases of macroscopic Bell non-locality. As
quantum states that demonstrate Bell non-locality involve strong
entanglement, the issue of preparing states for which
entanglement-destroying decoherence effects are minimised will be important,
since these effects tend to be more significant in macroscopic systems.
\medskip

\section{Acknowledgements}

The author wishes to acknowledge discussions with M. D. Reid, J. A. Vaccaro
and H. M. Wiseman, and thanks the referees for their helpful comments. BJD
also thanks the Centre for Cold Matter, Imperial College, London for its
hospitality during the writing of this article.\smallskip

\section{Contribution}

The paper was entirely written by the author. 

\end{document}